\documentclass[manuscript]{acmart}
\AtBeginDocument{%
  \providecommand\BibTeX{{%
    \normalfont B\kern-0.5em{\scshape i\kern-0.25em b}\kern-0.8em\TeX}}}

\usepackage{booktabs}
\usepackage{soul}
\usepackage{tabularx}
\usepackage{tabulary}
\usepackage[para,online,flushleft]{threeparttable}
\usepackage{tikz}
\usetikzlibrary{shapes.misc, positioning}
\usetikzlibrary{tikzmark}

\usepackage[frozencache,cachedir=.]{minted}

\usepackage{xcolor}
\definecolor{bg}{rgb}{0.9, 0.9, 0.9}
\newmintinline{python}{python3, fontsize=\normalsize, bgcolor=bg}
\usepackage{tablefootnote}
\usepackage{enumitem}
\usepackage{subcaption}
\usepackage{multirow}
\usepackage[capitalize]{cleveref}

\newcommand{\StudyParticipants}[0]{14}

\newcommand{\SurveyRespondents}[0]{114}

\begin{document}

\title{Conversational Challenges in AI-Powered Data Science: Obstacles, Needs, and Design Opportunities}\subtitle{}

\author{Bhavya Chopra}
\affiliation{%
  \institution{Microsoft}
  \country{India}}
\email{t-bhchopra@microsoft.com}

\author{Ananya Singha}
\affiliation{%
  \institution{Microsoft}
  \country{India}}
\email{t-asingha@microsoft.com}

\author{Anna Fariha}
\affiliation{%
  \institution{University of Utah}
  \country{United States}}
\email{afariha@cs.utah.edu}

\author{Sumit Gulwani}
\affiliation{%
  \institution{Microsoft}
  \country{United States}}
\email{sumitg@microsoft.com}

\author{Chris Parnin}
\affiliation{%
  \institution{Microsoft}
  \country{United States}}
\email{chrisparnin@microsoft.com}

\author{Ashish Tiwari}
\affiliation{%
  \institution{Microsoft}
  \country{United States}}
\email{astiwar@microsoft.com}

\author{Austin Z. Henley}
\affiliation{%
  \institution{Microsoft}
  \country{United States}}
\email{austinhenley@microsoft.com}

\renewcommand{\shortauthors}{Bhavya Chopra et al.}

\begin{abstract}
  Large Language Models (LLMs) are being increasingly employed in data science for tasks like data preprocessing and analytics. However, data scientists encounter substantial obstacles when conversing with LLM-powered chatbots and acting on their suggestions and answers. We conducted a mixed-methods study, including contextual observations, semi-structured interviews (n=14), and a survey (n=\SurveyRespondents{}), to identify these challenges. Our findings highlight key issues faced by data scientists, including contextual data retrieval, formulating prompts for complex tasks, adapting generated code to local environments, and refining prompts iteratively. Based on these insights, we propose actionable design recommendations, such as data brushing to support context selection, and inquisitive feedback loops to improve communications with AI-based assistants in data-science tools. 
\end{abstract}



\maketitle


\section{Introduction}

Data scientists have traditionally relied on resources like documentation, tutorials, online courses, colleagues, Q\&A forums, and online communities to develop skills and solve tasks~\cite{Kim2018TSE, Kross2019CHI, Shrestha:2021, vasilescu2014opensource}. 
These resources have been instrumental in helping them navigate the challenges of data acquisition, cleaning, wrangling, visualization, and presentation, especially for those with a non-programming background~\cite{kandel2012enterprise}.
Although useful, it can be time consuming, tedious, and error-prone to rely on these resources for solving a data-science problem.

With the emergence of AI-powered chat assistants, data scientists now have access to potentially faster and more accessible resources through a chat interface.
These AI-powered chatbots, like ChatGPT\footnote{https://chat.openai.com/}, enable users to ask questions in natural language, and get useful responses with little to no latency.
For example, when prompted with ``split my date column that is in MM/DD/YYYY format into three columns'', ChatGPT provided usable Python code:\\
\phantom{addingsomecenterit}\pythoninline{df[['Month', 'Day', 'Year']] = df['Date'].str.split('/', expand=True)}\\
ChatGPT can also explain how the code works, and supports follow-up questions or changes.
Evaluations of these AI tools have demonstrated over 70\% accuracy on programming benchmarks~\cite{Chen2021TR} and pass numerous tests designed for humans (e.g., the GRE, a college entrance exam, and the LSAT, a law school admission test)~\cite{OpenAI2023TR}.

However, the effectiveness of these tools is dependent on data scientists successfully communicating their questions, context (overall problem, task at hand, datasets, etc.), assumptions, and domain knowledge to the AI assistant, and doing so through a back-and-forth conversation.
Grice's maxims of conversation\footnote{https://effectiviology.com/principles-of-effective-communication/} posits that a successful conversation involves content that has right amount of information, that is truthful and supported by evidence, that is relevant to the specific context, and that is presented clearly~\cite{Grice1991Book}.
But conversations go awry---either side of the conversation may be making false assumptions, there may be ambiguities, and conversations may require numerous clarifications.
All of these problems are possible with AI assistants; most notably, they are known to \emph{hallucinate}, meaning they confidently respond with false statements. Furthermore, studies have found that users find conversations with AI assistants, such as ChatGPT, to have problematic communication styles~\cite{Skjuve:2023} and responses provided to be repetitive and filled with outdated or irrelevant information~\cite{Skjuve:2023, kaddour2023challenges}---a far cry from Grice's maxims.
As such, for humans to communicate effectively with AI agents, humans must follow an iterative sense-making process that involves describing intent, building hypotheses, gathering evidence, and evaluating responses~\cite{Cabrera2023TOCHI}.

The barriers to express intents are exacerbated in the context of data-science tasks.
First, data scientists work with a variety of artifacts, including raw datasets, code, computational notebooks, visualizations, documentation, and machine-learning pipelines. 
Second, datasets are often large and normalized (spilt across multiple tables), which may not be feasible to share (e.g., due to token limits of AI tools or due to data being spread across heterogeneous sources) or to summarize the relevant portions succinctly (e.g., specifying regions of the data). 
Third, real-world data is messy, and suffer from quality issues; they may not strictly adhere to a schema or homogeneous format.
Fourth, data-science tasks often require domain expertise and numerous assumptions (e.g., negative values in a column represent an error, 0 means missing value) and it may be laborious to ensure that the AI assistant is aware of such domain knowledge and assumptions. 
Furthermore, even when a data scientist is able to articulate their question to the AI assistant, these same problems can cause difficulties to understand and validate the AI assistant's response.

In this work, we aim to understand the fundamental problems of a [human] data scientist communicating to an AI chat agent. In particular, we address the following research questions:
\begin{itemize}
    \item \emph{RQ1:} How do data scientists interact with ChatGPT to complete data-science tasks?
    \item \emph{RQ2:} What challenges and unmet needs do data scientists face when interacting with ChatGPT?
    \item \emph{RQ3:} How well do these challenges and needs generalize to broader community of data scientists?
\end{itemize}

To answer these questions, we conducted two mixed-method need-finding studies (Section~\ref{sec:method}). For the first study, we observe \StudyParticipants{} professional data scientists as they perform four diverse tasks that are common in data-science pipelines. The participants were presented with the browser-based chat assistant, ChatGPT, to help them solve the tasks whenever needed. We performed inductive thematic analysis to identify needs and challenges by triangulating findings from open coding of semi-structured interview transcripts, observations from field notes, and video annotations. For the second study, we conducted a confirmation survey with \SurveyRespondents{} professional data scientists to validate and generalize findings from the first study.

Our results (Sections~\ref{sec:observations} and~\ref{Section:CSurvey}) show that data scientists usually follow a typical pattern of articulating their question, collecting context, validating the response made by the AI assistant, and then applying the response to their task. 
Throughout this process, they face copious barriers while communicating with the AI assistant, such as spending considerable time writing their question and collecting relevant context.
Then they face a variety of additional barriers while trying to make use of the AI assistant's response, such as understanding the assumptions being made by the AI and modifying the provided code to work for their scenario.
This arduous workflow resulted in participants applying strategies to overcome the barriers and still make use of the AI assistant. 
Based on our findings, we provide practical design recommendations that can improve user experiences while using LLM-powered assistants for their data-science workflows (Section~\ref{sec:design}).


\section{Background}\label{sec:background}

\subsubsection*{How data scientists work}
Data science is the broad discipline of using data to understand the underlying nature of a domain and provide insights~\cite{Piorkowski_CHI19}.
Activities include data acquisition, cleaning, wrangling, visualization, sharing, and acting on insights~\cite{Fisher2012Interactions, Kim2018TSE, Piorkowski_CHI19}.
The majority of time spent on data-science tasks often involves data cleaning and wrangling~\cite{Guo2011UIST, Kandel2011CHI}.  Sutton et al.\ stated that handling expansive datasets often involves navigating a myriad of issues, culminating in a situation they term ``death by a thousand wranglings''~\cite{Sutton2018KDD}.

To support exploratory wrangling tasks, data scientists often employ \emph{computational notebooks}, such as Jupyter notebooks\footnote{https://jupyter.org/}.
These notebooks provide an interactive environment that allows for the seamless integration of text, code, and output, facilitating both the development and documentation of computational workflows in a single document. 
Unlike traditional code editors, notebooks enable a more narrative style of coding, where code cells and rich-text commentary can coexist and outputs can be immediately rendered inline.
However, notebooks come with their own pain points: they take considerable effort to keep organized, hidden state information can lead to misconceptions, they can be difficult to deploy, and long-running tasks can be problematic~\cite{Kery2018CHI, Rule2018CHI, Titus_CHI20}.

\subsubsection*{Large language models and AI-powered chat assistants}
AI-powered chat assistants are built on top of large language models (LLMs).
LLMs are generative machine-learning models with billion parameters, and are trained on vast amount of data (text and images) to generate human-like responses and perform various language tasks.
These models interact with users through a \emph{prompt}, a natural-language query to which the LLM can respond to, also in natural language.
There are restrictions on the \emph{token limit}, or the amount of information that can be sent or generated from the LLM.
Tools like ChatGPT add a user interface to enable a conversation, such that the model has 
\emph{context} of the conversation history.
However, these models do not have direct access to data sources, and  only respond based on the data they have seen during training.


\section{Methodology}\label{sec:method}

We performed two studies to answer our research questions. First, we conducted task-based semi-structured interviews and observed \StudyParticipants{} professional data scientists engaged in data-science tasks. Then, we distributed a general survey of our initial findings with \SurveyRespondents{} data scientists.

\subsection{Study 1: Using ChatGPT for Data Science Tasks}
We used ChatGPT, a browser-based LLM-powered chat-based AI assistant, for the study. We avoided using any AI assistants integrated within computational notebooks to better understand fundamental challenges and patterns of usage. This deliberate choice to keep the AI assistant separate from the data science environment(s) of the participants allows us to give them complete control and transparency over the prompt-writing and code adaptation processes---while giving us an opportunity to understand the underlying process of assessing \textit{which} context, and \textit{how much} context is relevant for prompting. This choice also echoes one of the design recommendations proposed by \citet{Barke_OOPSLA23} that developers would prefer control over the shared prompt context.

\subsubsection*{Participants} We recruited participants from a large technology company by randomly selecting employees with the job title ``Data Scientist'' mentioned in the company address book. We further screened participants based on having at least 1 year of experience with data science and prior experience with Python, computational notebooks, and ChatGPT (or any other LLM-based chat/code assistants) as a pre-requisite to their participation. Table \ref{table:participants} indicates the demographics collected for all participants, who reported a median of 5 years of experience in the domain of data science ($\mu = 5.4$, $sd = 2.8$). All participants have a background and educational degree in CS and Data Science related fields.

All sessions were audio and video recorded for transcription of responses and analysis of on-screen interactions. All participants signed a consent form prior to the study. Participation in the study was voluntary and participants were compensated with gift cards worth \$25 USD each.

\begin{table}[h]
\caption{Overview of the participants, including the domain that they work in, their years of experience as a data scientist, and the programming languages they often use.}
{\small
\begin{tabular}{lllll}
\toprule
\multicolumn{1}{l}{\textbf{ID}} & \multicolumn{1}{l}{\textbf{Domain}} & \multicolumn{1}{l}{\textbf{Exp. (yrs)}} & \multicolumn{1}{l}{\textbf{Languages used}}  \\ 
\midrule
P1   & Anomaly Detection   & 5       & Python, SQL       \\
P2   & Big Data Processing                     & 10      & Python, SQL              \\
P3   & User Segmentation    & 4       & Python            \\
P4   & NL Processing             & 1       & Python                   \\
P5   & Feature Engineering   & 4       & SQL                      \\
P6   & Statistical Analysis                    & 2       & Python                   \\
P7   & Defect Detection         & 5       & Python                   \\
P8   & Finance \& Forecasting                  & 6       & Python, SQL, Spark       \\
P9   & Business Analytics                      & 2       & Python                   \\
P10  & Fraud Detection                         & 5       & Python, SQL           \\
P11  & User Segmentation                       & 10      & Python, SQL, Spark       \\
P12  & Cloud Computing                         & 10      & Python                   \\
P13  & Failure Prediction                      & 7       & Python                   \\
P14  & Fraud Detection                         & 5       & Python, SQL       \\ 
\bottomrule
\end{tabular}
}
\label{table:participants}
\end{table}

\subsubsection{Tasks}

We used four essential tasks performed by data scientists in the lifetime of a typical project~\cite{Piorkowski_CHI19}: (T1 and T2) data wrangling and pre-processing, (T3) feature extraction and selection for training tasks, (T4) data visualization for insight-finding and reporting. The four tasks are described in Table~\ref{table:tasks}.

We selected a sample of the \emph{New York EMS emergency calls dataset}\footnote{\url{https://www.kaggle.com/datasets/new-york-city/ny-ems-incident-dispatch-data}}, which contains rows corresponding to emergency incidents, and information about time, location, resources, and the Fire Department’s response to the emergency. The dataset was well-suited for supporting the tasks, as several columns had to be transformed before performing further analysis, and event data was amendable to feature and correlation analysis.

\begin{table}[t]
\caption{The four data science tasks used in our study and their descriptions.}
{\small
\begin{tabular}{p{0.3cm} p{2.5cm} p{11cm}}
\toprule
\multicolumn{1}{c}{\textbf{\#}} & \multicolumn{1}{c}{\textbf{Task}} & \multicolumn{1}{c}{\textbf{Description}} 
\\ \midrule
T1  & Datetime typecast    & Change the datatype of column \texttt{``INCIDENT\_DATETIME''} from string to datetime (e.g., \texttt{``2020-07-31T23:59:51.000''}).\\\addlinespace

T2  & Split using delimiter  & Split the (\texttt{``INCIDENT\_DESCRIPTION''}) column using semicolon (\texttt{``;''}) as the delimiter. Rows contain information about the incident location --- borough, incident dispatch area code, and zipcode. Few rows do not contain the zipcode (e.g. rows such as \texttt{``Brooklyn; K7; 11211''} contain all information, whereas rows like ``Manhattan; M3'' are missing the zipcode).\\\addlinespace

T3  & Feature selection  & Extract a new column titled \texttt{``RESPONSE\_TIME''} using the difference between \texttt{``FIRST\_ON\_SCENE\_DATETIME''} and \texttt{``INCIDENT\_DATETIME''}; subsequently perform feature selection with the aim of predicting the \texttt{``RESPONSE\_TIME''}.\\\addlinespace

T4  & Heatmap plotting  & Obtain a heatmap plot to analyze correlation between any columns of choice.\\
\bottomrule
\end{tabular}
}
\label{table:tasks}
\end{table}

\subsubsection*{Study protocol} We conducted a one-hour long, task-based study with each of the \StudyParticipants{} data scientists via video conferencing. The study begins with demographic questions about their role and data-science-related work, languages and tools, educational background, and prior experience with using LLMs. Next, participants were presented with 4 tasks, one at a time, to be solved using Python in any computational notebook of their choice, and the browser-based ChatGPT\footnote{\url{https://chat.openai.com/} (Versions: March 14, 2023; March 23, 2023)}. They were encouraged to make use of any preferred tools and resources (such as web search, notebooks, Excel, and so on), and think aloud while solving the tasks. The completion of any task was not a necessity, and participants were encouraged to use ChatGPT at least once for every task. Once the participants had explored a task sufficiently, they were asked semi-structured questions to assess if ChatGPT helped them solve the task satisfactorily, and if it could have helped them with the task any better.

\subsubsection*{Analysis} We transcribed the audio recordings for each session. Subsequently, we annotated videos from each session recording to timestamp and gathered: (1)~prompts authored by participants, (2)~corresponding responses provided by ChatGPT, and (3)~a log of activities performed by the participants (including prompt writing, adapting generated code, and validating generated code).
The first and second author analyzed the transcripts and annotations from video recordings to perform inductive thematic coding \cite{saldana-2009}. We first conducted discussions over a span of 2 weeks while collaboratively annotating data for one participant. Following this, we independently annotated 20\% of the data (from three participants) with descriptive open codes. Next, we grouped similar codes and analyzed them to obtain axial codes, identifying usage patterns and pain-points faced by data scientists. We then computed the inter-rater agreement score to be 0.872, indicating `almost perfect' agreement \cite{Landis_Koch_1977}. The first two authors then split the remaining participant interviews and  annotated them independently, while meeting frequently to discuss and refine the codes and themes.

We observed theoretical saturation after thematic analysis of data from 9 participants. To further support the validity of our findings, we requested 5 study participants (P1, P3, P4, P8, P12) to respond to a member-check survey. The survey consisted of 18 Likert-scale questions, each question representing an open-code from our findings. The ratings indicated agreement with each finding.

\subsection{Study 2: Confirmatory Survey}
To validate findings from Study 1, we conducted a survey with a broader population of \SurveyRespondents{} data scientists at a large technology company.

\subsubsection*{Survey protocol} The survey consisted of two demographic questions, asking respondents to: (D1)~indicate years of experience with data science (a single choice question), and (D2)~describe at least one scenario where they used any LLM for a data-science task (an open-field question). To ensure the quality of the survey responses, we used responses for D2 to filter out respondents who do not have any experience with solving data-science tasks using LLMs.

We used findings from the qualitative analysis of Study 1 responses to frame at least one question for each identified obstacle. The survey consisted of 8 `Agree'/`Disagree' statements, and all questions were optional to answer. Finally, to evaluate if our observations from Study 1 reached theoretical saturation, we asked respondents to indicate if they have faced any other difficulties in using LLMs for data science tasks that were not mentioned in the survey questions.

\subsubsection*{Recruitment and informed consent} The respondents were contacted randomly via e-mail invites based on their job title (``Data Scientist'') mentioned in the company address book. In our e-mail invites, we also requested participants to forward the survey to other data scientists who might be interested. We offered a drawing for one of two Amazon gift cards worth \$50 USD each for completing the survey.

\subsubsection*{Respondents} 114 data scientists from a large technology company responded to the survey. After inspecting whether the respondents have meaningful prior experiences (D2), we removed 16 respondents ($14\%$) that had no experience with using LLMs for data science tasks. We report all findings using the screened responses. $76\%$ of the respondents self-reported having more than 5 years of experience in data science related professional roles. We discuss survey results in Section \ref{Section:CSurvey}, where we plotted the responses and inspected them manually to gather any other difficulties that were not mentioned in the survey questions.

\begin{figure}[t]
        \centering
        \includegraphics[width=\linewidth]{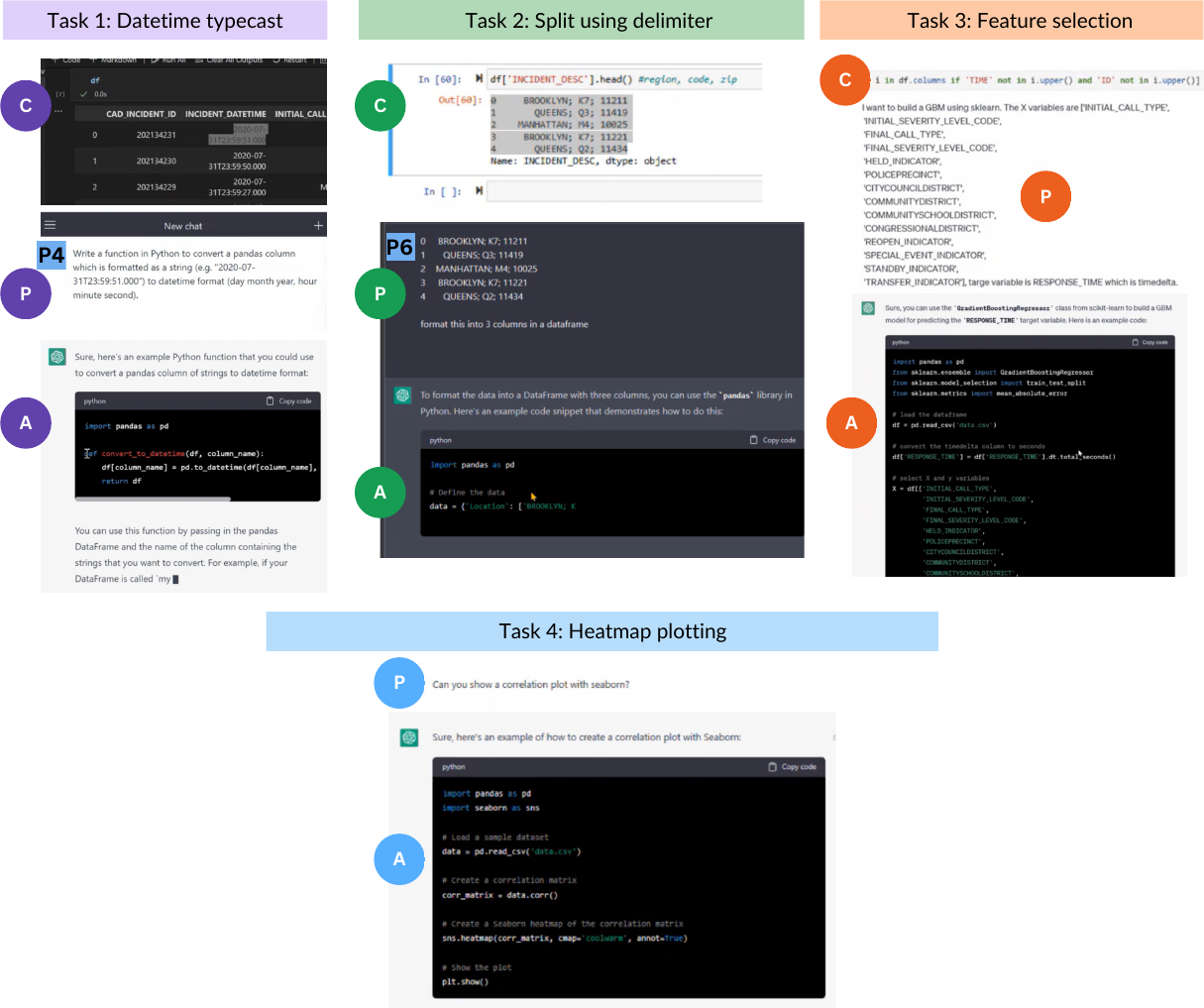}\
        \caption{A gallery of tasks involved in Study 1. Images are taken from participant screen sharing recordings and are anonymized. Snippets marked (C) denote the local \textit{\textbf{Context}} retrieved by participants to use in prompt-writing; (P) denotes the \textit{\textbf{Prompt}} sent to ChatGPT; and (A) denotes the \textbf{\textit{Answer}} provided by ChatGPT.}
        \label{fig:gallery}
    \end{figure}


\section{Observations}
\label{sec:observations}

We provide a summary of our participants interactions with ChatGPT and then report observed obstacles and strategies used by the participants.
Table~\ref{table:themes} summarizes the themes that emerged from our analysis of the data scientists' actions and responses. We also report participants who identified with each obstacle and employed prompting strategies in Table~\ref{table:themes}.

\subsection{Participants Interactions with ChatGPT}

\subsubsection{Typical workflow}

Most participants started with loading the CSV files into a notebook of their choice, and constructing a \texttt{pandas} dataframe, while viewing the first few rows of the data. Sometimes they opened the CSV in Excel before creating a notebook, to glance over the data. Some participants used \pythoninline{pandas.DataFrame.describe()} to obtain a high-level summary of their data. 

Next, they were sequentially presented with each of the tasks. Every participant worked on tasks T1, T2, and T3. However, 4 participants were unable to attempt T4 in the available time. For each task, participants generally began with phrasing their question, assessing and gathering context---such as description of the data, sample values---for prompt construction, understanding and taking action on ChatGPT's response while writing additional prompts that would help them elicit better responses. Figure \ref{fig:gallery} presents a gallery of tasks involved in the study. 

Participants faced several challenges in conversing with ChatGPT and using its responses for task completion. In particular, most data scientists struggled with T3 (feature selection), taking them significantly higher number of prompts and time to get a satisfactory response. Feature selection, being a broad task (as compared to the remaining study tasks), had participants thinking about decomposing it to plan for sub-tasks, and providing ChatGPT sufficient context to obtain concrete and actionable next steps.

\subsubsection{Prompting behaviors}\label{sec:prompt_behavior}

\begin{figure}[t]
    \begin{minipage}{0.45\linewidth}
        \includegraphics[width=\linewidth]{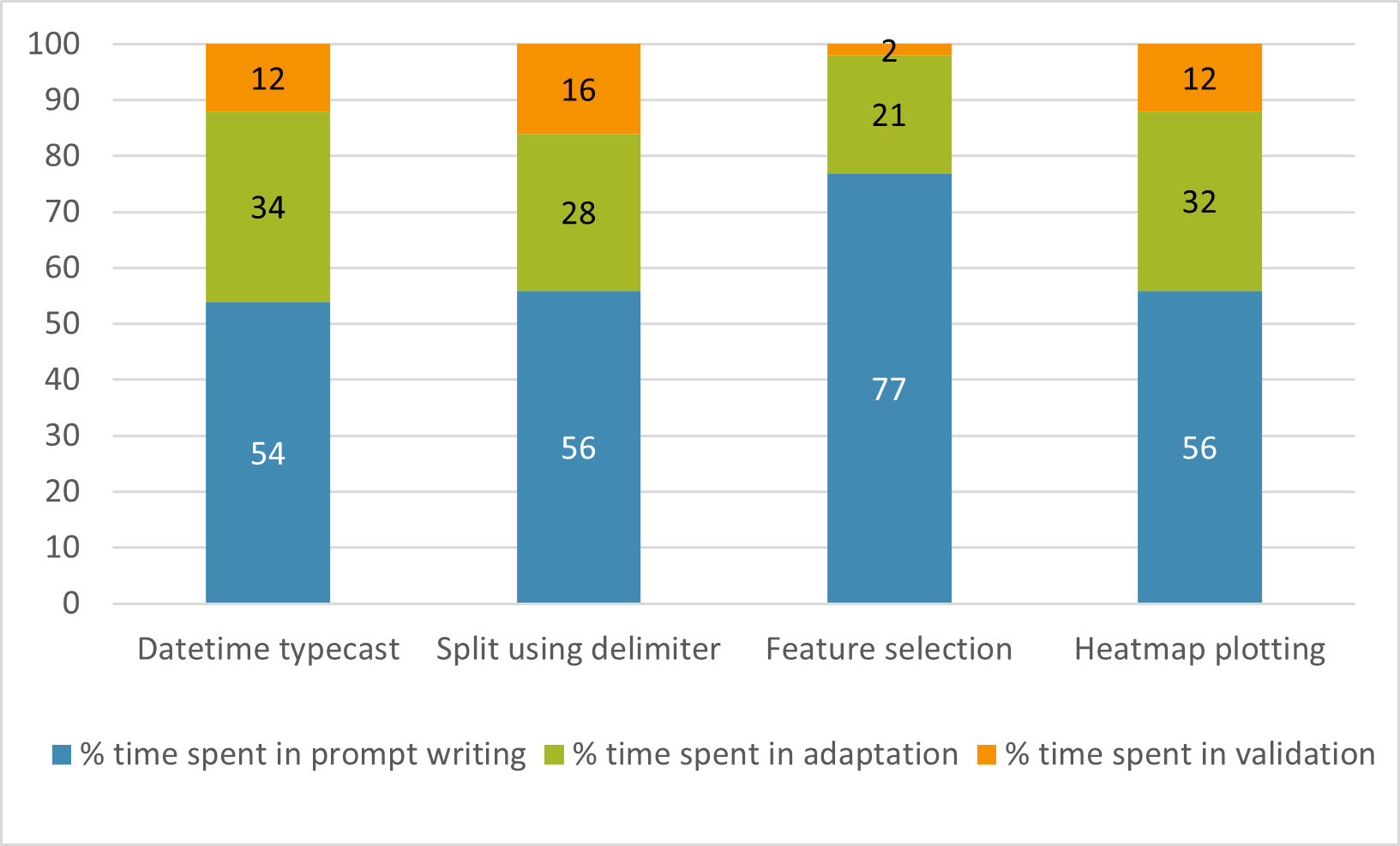}
    \end{minipage}%
    \quad
    \begin{minipage}{0.45\linewidth}
        \includegraphics[width=\linewidth]{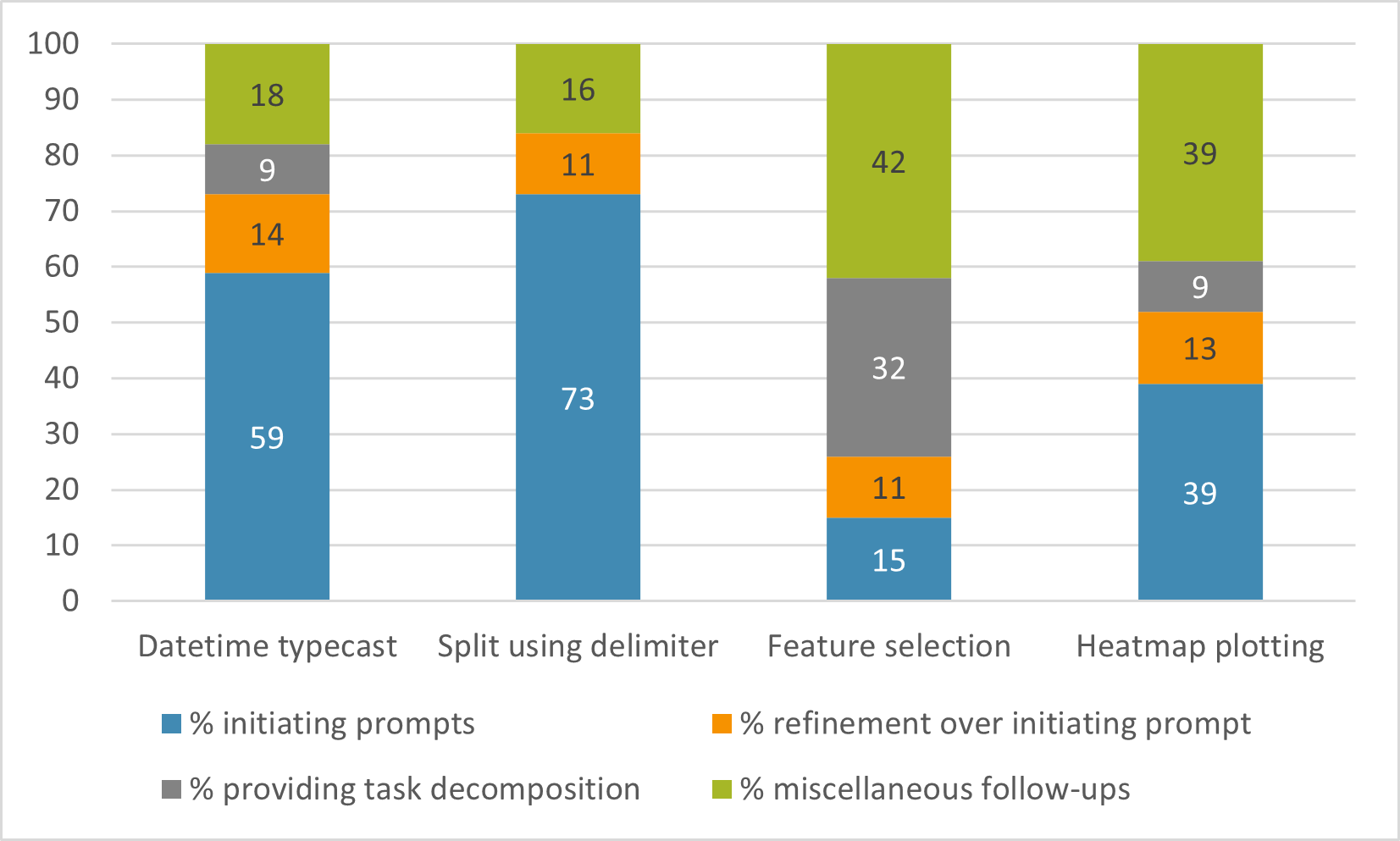}
    \end{minipage}
    \caption{Percentage of total time spent in each activity --- prompt writing, code adaptation, and validation --- for each task (Left).
    Stacked percentage distribution of initiating prompts, refinements, prompts for task decompositions, and follow-up prompts (Right).}
    \label{fig:time-spent-percentage}
    \label{fig:prompt-type-percentage}
\end{figure}

In total, we logged 111 prompts made by \StudyParticipants{} participants. Table~\ref{table:prompts-stats} presents an overview of the prompt distribution for each study task. We observed that participants spent $64\%$ time on preparing prompts, $27\%$ time on adapting the code returned by ChatGPT, and the remaining on validating the code, as seen in Figure \ref{fig:time-spent-percentage} (Left). In raw time spent on these activities, on average, $302$ seconds were spent on prompt writing, $57$ seconds on adaptation and $24$ seconds on validation. We also find that $37\%$ of the total time was spent on writing the initial prompt, whereas making refinements to prompts and asking follow-up questions takes $28\%$ of the total time.
In particular, we observed that participants especially struggled to write an initial prompt for T3 (feature selection), a task with more subjectivity, when compared to relatively more objective tasks (T1, T2 and T4). The prompts also greatly varied in verbosity:

\begin{quote}
\itshape
You are a data scientist, how can you do feature selection? (P13)
\end{quote}

\begin{quote}
\itshape
I have a dataset that has information about medical emergency services. I have a response time column which tells me historically how much time will it take to respond to a medical emergency incident. I have other columns some are categorical and some are continuous which explain more about the incident. My task is to predict the response time for a new incident and I am thinking about building a predictive model for that. What would be good features that can help the model predict response time? (P8)
\end{quote}

\begin{table}[t]
\caption{Overview of the number of prompts queried per task, and average time spent in prompt-writing.}
{\small
\begin{tabular}{lllllll}
\toprule
\textbf{Task}         & \multicolumn{1}{c}{\textbf{\# prompts}} & \textbf{median} & \multicolumn{1}{c}{\textbf{min}} & \textbf{max} & \textbf{avg time spent (s)} \\ 
\midrule
T1 -- Datetime typecast     & 22          & 1     & 1     & 3    & 55    \\
T2 -- Split using delimiter & 19          & 1     & 1     & 2    & 59    \\
T3 -- Feature selection     & 47          & 3     & 0     & 6    & 159   \\
T4 -- Heatmap plotting      & 23          & 2     & 0     & 4    & 58    \\
\bottomrule
\end{tabular}}
\label{table:prompts-stats}
\end{table}
      
Participants also made iterative refinements to their prompts. They sent across a plenitude of follow-up prompts for performing feature selection (T3)---re-emphasizing task goals, decomposing tasks, providing additional context and examples, and requesting for tweaks to code snippets. The need to send follow-up prompts was also exacerbated for plotting visualizations (T4), where participants would go back-and-forth with ChatGPT for minor tweaks---changes in plot size, color scheme, and font size. Figure \ref{fig:prompt-type-percentage} (Right) reflects on the distribution of follow-up and refined prompts for each task.

Overall, we observed that when participants interacted with ChatGPT as an assistant for providing recommendations for code, it often involved many steps for more involved tasks. Further, participants often struggled with providing information about their data and desired goals to ChatGPT and adapt its responses to solve their tasks.

\subsection{Obstacles when Communicating with ChatGPT}
\label{obstacles-communicating}

    \subsubsection{Sharing context is difficult}
    \label{context-sharing}

        \textbf{\textit{What information must be shared with ChatGPT?}} Since LLMs \emph{``cannot make sense of the raw data''} (P2, P3, P7, P8, P10, P12), data scientists begin the process of prompt-writing by brainstorming what pieces of data, (such as filename, column names, column data types, sample rows, etc.) and which descriptive information (such as pattern of strings in a column, number of unique values, missing values, outlier information, minimum/maximum for numeric data, etc.) will help ChatGPT in solving the query satisfactorily. Participants reported that starting with curating a prompt is a ``daunting'' task (P9) because it requires decision-making on what data context needs to be included. P2 said, \emph{`` I doubt if this will be successful, I feel like we need a little bit more of numbers, but let's see if only using the column names we can get something.''}

        \begin{table}[t]
\caption{Themes---obstacles and strategies when using ChatGPT for data-science tasks.}
\scalebox{0.92}{
\small
\begin{threeparttable}
\begin{tabularx}{\textwidth}{>{\raggedright}p{3cm}p{4cm}Xp{2cm}}
  \toprule
\textsc{Theme} & \textsc{Description} & \textsc{Representative Examples} & \textsc{Participants}\\

\midrule
  \multicolumn{3}{l}{\textbf{Obstacles when Communicating with ChatGPT}}\\
\midrule
  
  \emph{Sharing context is difficult}\newline{\small (Section~\ref{context-sharing})}
  & Gathering relevant information to prompt with (e.g., column names, datatypes, example datapoints) takes time.
  & ``Should I add some rows for it to see? I am not sure if [it] will work...'' \newline ``So this one is recommending \texttt{scikit-learn} which is not a package I typically use. I usually work in \texttt{pyspark}.'' \newline ``It is all about the context which we provide. If we could refine it, it would do better.'' & P1---P6, P8, P10, P12, P14\newline (10 \textit{of} 14)\\\addlinespace
  
  \emph{ChatGPT opaquely makes assumptions}\newline{\small (Section~\ref{assumptions})} 
  & Data scientists had to correct or adjust prompts in response to unanticipated assumptions made by ChatGPT about data.
  & ``I will provide this example for the format'' \newline
  ``Oh, it thinks the call type is int, let me say that it is categorical''
  & P1---P14\newline (14 \textit{of} 14)\\\addlinespace
  
  \emph{Misaligned expectations}\newline{\small (Section~\ref{misaligned-expectations})}
  & ChatGPT often returned overly verbose answers or inappropriate code responses.
  &  ``I will skip reading any of this text, unless my code doesn't work'' \newline ``it didn't bring up \texttt{to\_datetime()} and suggested \texttt{strptime()} instead!''  & P1---P10, P12---P14\newline (13 \textit{of} 14)\\\addlinespace

  \midrule
    \multicolumn{3}{l}{\textbf{Obstacles in Leveraging ChatGPT's Responses}}\\
 \midrule

  \emph{Generation of repeated code}\newline{\small (\cref{repeated-steps})} 
  & ChatGPT repeatedly generated code for imports, data ingestion and exploration, which had to be removed.
  & ``It erased my data, I will quickly re-run the cells'' \newline
  ``We already have pandas and the data, I'll delete these lines [...] not required''
  & P1---P10, P12---P14\newline (13 \textit{of} 14) \\\addlinespace

   \emph{Data and notebook management preferences}\newline{\small (\cref{code-mgmt-preferences})} 
  & Splitting code into cells and aligning it with data management preferences (such as creating temporary views, dropping parent columns, etc).
  &  ``a temp df to see the new split columns'' \newline
  ``I would not fill with ‘NULL’ [...] still need to understand these columns more''
  & P2---P9, P12---P14\newline (11 \textit{of} 14)\\\addlinespace
  
  \emph{Code validation}\newline{\small (\cref{code-validation})} 
  & Data scientists often author additional code snippets to validate generated code.
  &  ``ChatGPT's affirmative language is deceiving [...] it doesn't really know my data'' \newline
  ``let me print \texttt{df.dtypes} to be sure''
  & P1---P10, P12---P14\newline (13 \textit{of} 14) \\\addlinespace

   \midrule
    \multicolumn{3}{l}{\textbf{Strategies for Prompting and Alternate Resources}}\\
 \midrule
  
  \emph{Techniques for prompt construction}\newline{\small (\cref{placeholders})} 
  & Data scientists use prompting techniques and create placeholders for streamlined context gathering from various sources 
  & ``lets ask why it selected these features, why not the others'' \newline
  ``I will paste column names later, so that I can write the prompt first''
  & P1---P10, P12, P14\newline (11 \textit{of} 14) \\\addlinespace

  \emph{Scaffolding with domain expertise}\newline{\small (\cref{scaffolding-domain-expertise})} 
  & Data scientists often employ techniques to guide ChatGPT with their domain knowledge to elicit better responses.
  & ``I can be more specific and give it only the float columns''\newline
  ``I will exclude the time columns to get good features''
  & P5, P10, P14\newline (3 \textit{of} 14) \\\addlinespace

  \emph{Choosing alternate resources over ChatGPT}\newline{\small (\cref{alternatives})} 
  & Making use of web-search; and re-use of previously authored notebooks/pipelines 
  & ``must write longer queries instead of 8-10 key words just because I can''
  \newline ``copy paste canned code [...] 80\% of it would be repeating''
  & P3---P5, P7, P11, P12\newline (6 \textit{of} 14) \\\addlinespace
  
  \bottomrule
\end{tabularx}
\end{threeparttable}}
\label{table:themes}
\end{table}

        \textbf{\textit{How much information must be shared with ChatGPT?}} Participants often questioned \emph{how much} information needs to be shared with ChatGPT to get a good response. Some participants made attempts to paste the entire CSV file into the prompt and encountered token limit errors (P6, P7, P13). Others also reflected on the importance of sharing only relevant instructions or snippets of the data as \emph{``additional data seems to confuse ChatGPT into providing not-so-relevant results''} (P7, P9, P10). P9 said, \emph{``The main obstacle I ran into here was, knowing how to prompt ChatGPT. Because I think at first I almost gave it too much information on how to perform a correlation analysis by describing it.''} After sharing 10 sample rows, P2 tried to estimate ChatGPT's boundaries in being able to understand raw data beyond its semantics, as it provided a \emph{``very basic response for the most relevant fields, by repeating the entire fields names and telling me their [semantic] meaning.''}

        \textbf{\textit{Challenges in fetching context.}} To construct the prompt, after assessing \textit{what} context is relevant, participants had to go back and forth between ChatGPT's browser window and data sources to fetch the context. They had to frequently write code snippets in their computational notebooks to gather the required context. For instance, consider a \texttt{pandas} DataFrame named \texttt{df}. Participants must write, execute, and copy the output of commands such as: 
        \begin{itemize}
            \item \pythoninline{df.head()} to obtain a sample of rows and the data header
            \item \pythoninline{df.columns} or \pythoninline{df.dtypes} to obtain column names and their data types
            \item \pythoninline{df['INCIDENT_DATETIME'][0]} and \pythoninline{df['INCIDENT_DESCRIPTION'][0]} to obtain the first value in the respective columns to provide as few-shot examples, and so on.
        \end{itemize} 

        We observed that P10 and P14 also implemented custom logic to extract information to share beyond what is typically shared in a prompt. For instance, P14 wanted to provide names of specific columns to ChatGPT and hence wrote the following code that computes column names that do not contain the sub-strings \texttt{`TIME'} or \texttt{`ID'}:

    \begin{minted}[escapeinside=||,fontsize=\small,breaklines]{python}
        >>> [i for i in df.columns if 'TIME' not in i.upper() or 'ID' not in i.upper()]
    \end{minted}
        
        \begin{figure}[t]
            \centering \includegraphics[width=\linewidth*8/10]{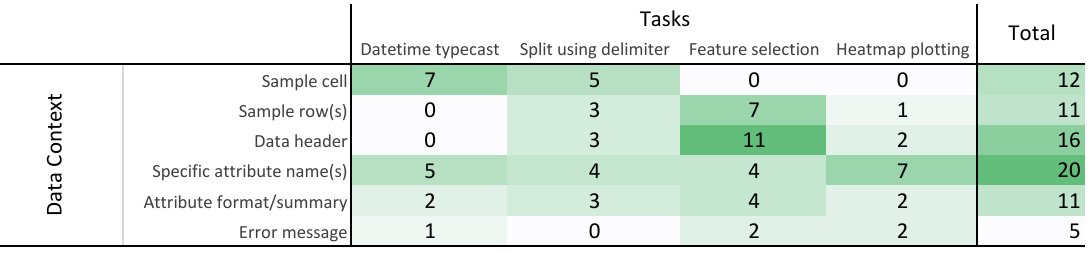}\
            \caption{Frequency of data context shared by participants for prompt-construction for each task.}
            \label{fig:freq_context}
        \end{figure}
        
        Further, P5, P7 and P8 had to leave their notebook environment to open the CSV file in Excel. This was because the participants wanted to share values from a column, and the raw output from \pythoninline{df.head()} could not be selected freely to be copied to the clipboard. Opening the file in Excel enabled them to select any range of cells and copy them with ease without writing additional code.

        In addition to data context, participants also used snippets of their code with error messages to prompt ChatGPT for a fix (P6, P7, P12). Figure \ref{fig:freq_context} shows the frequency of different types of data context shared with ChatGPT corresponding to each task.

        After fetching the relevant data context, participants had to paste it into the prompt. Participants would often manually format the pasted information to make the prompt look \emph{``cleaner,''} to ensure that ChatGPT can understand the context (P2, P3, P7). While solving T3 (feature selection), P2 wanted to fetch column names, and organize them such that each column name appears in a new line in the prompt. To this, he said, \emph{``Can you copy and paste it for me please? By any chance would you have like all the columns of the dataset one name per line? In the meantime, I'll be typing [the prompt]... I am so lazy.''}

    \subsubsection{ChatGPT opaquely makes assumptions}
    \label{assumptions}
         Based on the context shared with ChatGPT, it made its own ``mental model'' of the data. Participants were often enthusiastic about the semantic capabilities of ChatGPT, and how it could infer domain knowledge about the data solely using the column names. P4 leveraged these capabilities and prompted ChatGPT to provide a data dictionary. Several participants (P5, P7, P10, P12---P14) were amazed when ChatGPT could understand the data domain sufficiently to convert the extracted \texttt{`RESPONSE\_TIME'} column from \texttt{timedelta} format to seconds---which was the most appropriate unit of measurement, since emergency medical services are likely to be dispatched quickly. However, despite the semantic abilities of the LLM, participants struggled with several scenarios, and had to re-assess ChatGPT's understanding of the task and the data. We observed several instances across every task, where lack of adequate context in the prompt led ChatGPT to make assumptions about the data and its format. 
         
         Several participants (P1, P5, P6) received incorrect code generations for T1 (datetime typecast), where ChatGPT assumed the string to be of the format \texttt{`\%d\%m\%Y \%H:\%M:\%S'} (e.g., 31-07-2020 23:59:51) instead of \texttt{`\%Y-\%m-\%dT\%H:\%M:\%S.\%f'} (e.g., 2020-07-31T23:59:51.000). This assumption led to generation of the following code snippet to typecast the string column to datetime format, resulting in a run-time error: 
        \begin{minted}[escapeinside=||,fontsize=\small,breaklines]{python}
        >>> pd.to_datetime(date_str, format='%d%m%Y %H:%M:%S')
        Error: time data '2020-07-31T23:59:51.000' does not match the format '%d%m%Y %H:%M:%S' (match) 
        \end{minted}
        Upon providing ChatGPT with the error message, or an example string from the column, it re-generated the snippet without the \texttt{format} parameter, allowing successful execution, as \texttt{pandas} can now correctly infer the format of the input datetime string itself. 

        \begin{sloppypar}
        Next, as part of T2 (split using delimiter), participants had to work with the \texttt{`INCIDENT\_DESCRIPTION'} column, containing the borough, incident dispatch area code, and zipcode, separated by semicolons. However, few rows in this column did not contain the zipcode. This led to the column having two formats: (1) \texttt{``<borough>; <incident\_dispatch\_area\_code>; <zipcode>''}, for example: ``Brooklyn; K7; 11211'', and (2) \texttt{``<borough>;} \\\texttt{<incident\_dispatch\_area\_code>''}, for example: ``Manhattan; M3''. Participants generally did not discover this data-formatting inconsistency by themselves, since encountering such issues requires exploratory data analysis and on-ramping. As a result, ChatGPT often did not get visibility into this data quality issue through participant prompts (P1, P2, P5, P11), and it generated non-executable code for splitting the column under the assumption of having clean and consistent data, which contains all three splits in each row:
        \end{sloppypar}
        \begin{minted}[escapeinside=||,fontsize=\small,breaklines]{python}
        # split the INCIDENT_DESCRIPTION column by the ; separator
        >>> df['Borough'] = df['INCIDENT_DESCRIPTION'].str.split(';').str[0]
        >>> df['Precinct'] = df['INCIDENT_DESCRIPTION'].str.split(';').str[1]
        >>> df['ZipCode'] = df['INCIDENT_DESCRIPTION'].str.split(';').str[2]
        IndexError: list index out of range
        \end{minted}

        \vspace{1em}
        
        On the other hand, some participants were able to receive correct code generations from ChatGPT owing to their relatively generic prompts. In such cases, where the generated code ran without errors, participants never discovered the underlying data quality issue---leading them to be unaware about missing values in the newly derived column for zipcode (P1, P3--P4, P7--P14). P4 expected similar behavior for other data smells and quality issues, such as \emph{``typos and capitalization inconsistencies in categorical variable columns,''} and expressed the \emph{need} to be able to quickly discover and convey such data quality issues to the LLM.
        
\begin{figure}[h]
            \centering \includegraphics[width=\linewidth]{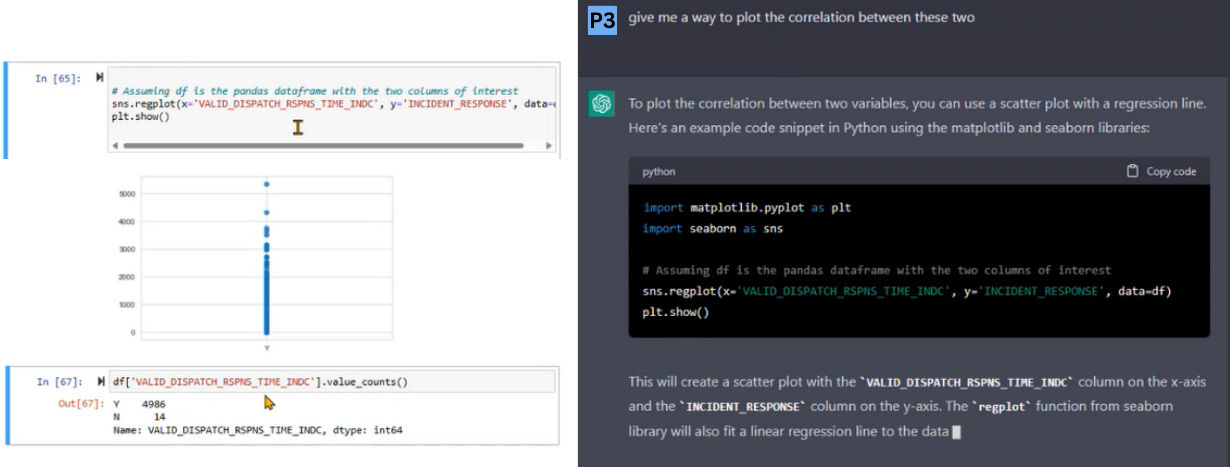}\
            \caption{ChatGPT suggests a highly skewed boolean variable for a scatter plot. P3 says that, \emph{``Should I choose another column? [...] This being categorical and skewed isn't useful for the visualization. So this is where I think [ChatGPT] only looked at the column names, and not the actual values in the data I shared.''}}
            \label{fig:t4-assumption}
        \end{figure}

        In tasks T3 (feature selection) and T4 (heatmap plotting), ChatGPT often made assumptions about the data type of columns solely based on the column names. That is, when participants did not share sample rows, or the column data types, ChatGPT used the column names to make a semi-informed guess about the data type. This led to run-time errors in T3 while trying to train a model (P5, P12), and visualizations that were not meaningful for any insight-finding in T4, such as scatter-plots for categorical and boolean variables (seen in Figure \ref{fig:t4-assumption}) (P2, P3, P10).

    \subsubsection{Misaligned expectations}
    \label{misaligned-expectations}

        Responses generated by ChatGPT are highly temperamental to the phrasing of prompts. Participants form a mental model of the nature of responses they would get from ChatGPT, in terms of whether the response will contain code or not, and the libraries, packages and APIs it would use in code generation. However, participants report that the presence or absence of specific words in the prompt can lead to unanticipated responses from ChatGPT. 
        
        For instance, not mentioning ``pandas'' or ``dataframe'' in the initial prompt to ChatGPT led to code generations with APIs belonging to standard python libraries, instead of \texttt{pandas} APIs (P1, P3, P12). P12 reflected on the issue and realized that they \emph{``didn't talk about pandas, that's why it didn't bring up \pythoninline{to_datetime()} over here and suggested \pythoninline{strptime()} instead!''}

        \begin{quote}
        \itshape
        Convert datetime string like ``2020-07-31T23:59:51.000'' to datetime object in python (P12)
        \end{quote}
        \begin{minted}[escapeinside=||,fontsize=\small,breaklines]{python}
        >>> from datetime import datetime
        >>> datetime_str = '2020-07-31T23:59:51.000'
        >>> datetime_obj = datetime.strptime(datetime_str, '%Y-%m-%dT%H:%M:%S.%f')
        \end{minted}

        Similar challenges surfaced as participants solved T3 (feature selection). Owing to the subjectivity of T3, participants often mentioned that ChatGPT cannot be completely trusted for suggesting techniques and steps since it lacks domain expertise (P1--P5, P7, P8, P11). Participants observe that ChatGPT's responses are almost never aligned with their domain knowledge, as it misses necessary data pre-processing steps, or provides technically incorrect responses. In one instance, as P5 went about prompting to train a model that predicts the \texttt{`RESPONSE\_TIME'}, ChatGPT generates code to perform training without standardizing column values. P5 emphasized the importance of standardization and suggested careful examination of all generated code: \emph{``One thing it has missed is standardizing data. It is very important for us to scale the data to check for outliers and understand the data distribution.''} In other instances, ChatGPT suggested selecting the features: \texttt{`INCIDENT\_DATETIME'} and \texttt{`FIRST\_ON\_SCENE\_DATETIME'}, which were used to compute the \texttt{`RESPONSE\_TIME'} column. Participants expressed that this is technically incorrect since the target column was arithmetically derived using the features selected by ChatGPT (P2, P3, P5, P13, P14); and that the \texttt{`FIRST\_ON\_SCENE\_DATETIME'} information will not be available in a pragmatic setting---making it impossible to predict the \texttt{`RESPONSE\_TIME'} (P8).

        Lastly, participants often found ChatGPT's responses to be \emph{``excessively verbose and unnecessary''}, especially for natural language (NL) code explanations. Participants wanted ChatGPT to omit explanations for APIs that they were already familiar with, and required NL explanations for only those unknown or uncommon APIs and parameters. Several participants held the long NL explanations for code generations as responsible for the prolonged inference time, that is, the time taken by ChatGPT to produce a complete response. P7 said, \emph{``When it was generating the explanation content, I felt like it is time consuming. Just give me the code. I was being impatient.''}       
    
        We observed that all participants skip reading any NL explanations when the response included a code-snippet, and begin with copying the generated code to adapt it to their notebook. P4 said, \emph{``I guess it takes a bit of time to finish running. It would just be easier for me if it just outputs the code, and then I can ask follow up question on what does $n = 2$ in the first line of the function do.''}

\subsection{Obstacles in Leveraging ChatGPT's Responses}
\label{obstacles-adapt-code}

\begin{figure}[t]
        \centering
        \includegraphics[width=\linewidth*8/10]{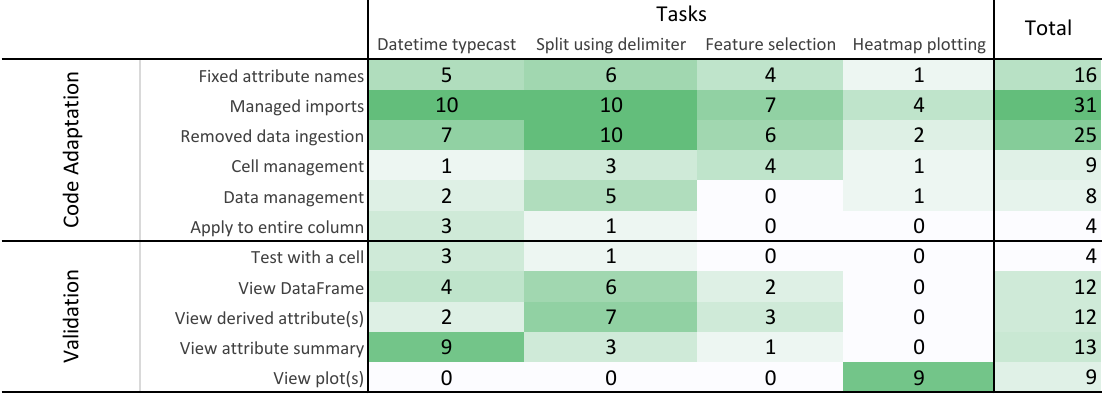}\
        \caption{Frequency of code adaptation and validation activities performed by participants for each task.}
        \label{fig:freq_adaptation}
    \end{figure}

    We now present the challenges faced by participant in {\em{using}} the code generated by ChatGPT. Participants had to edit the generated code in certain ways to successfully adapt it to their notebook contexts. Since the chat instance with ChatGPT was completely detached from the notebook, unless participants explicitly provided variable and attribute names in their prompt, ChatGPT created dummy dataframes and attribute names and included them in the generated code. Thus, the adaptation process frequently involved copying the generated code and replacing the generated variable and attribute names with the actual identifiers. These trivial, yet additional, steps in the adaptation process can be fixed by augmenting the user prompt (i.e., scaffolding the user's NL prompt with a prefix, and/or suffix to provide relevant context to the model) with the name of the dataframe, attribute names, and any other variables involved. However, we identify several other repeated obstacles across participants, which either require more involved prompt augmentation, post-processing of the generated code, or interface design interventions for easier adaptation. Figure \ref{fig:freq_adaptation} details the nature of code adaptation and validation activities and their frequencies across tasks, with managing import statements and removing additional code generated for data ingestion being the most frequent adaptations.

    \subsubsection{Generation of repeated code}
    \label{repeated-steps}
\begin{figure}[h]
            \centering \includegraphics[width=\linewidth*9/10]{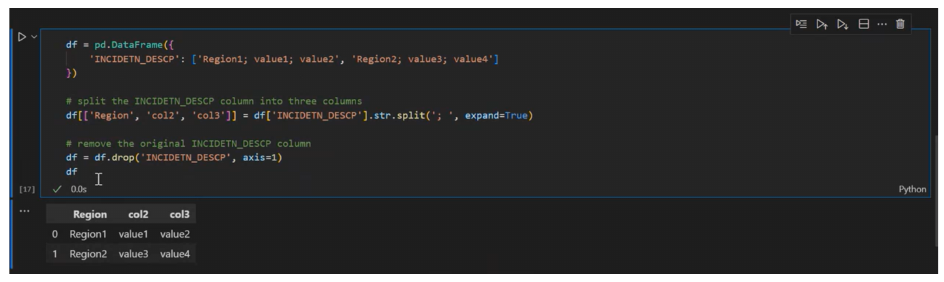}\
            \caption{P1 accidentally executes data ingestion code generated by ChatGPT, causing \texttt{df} to be over-written by a dummy dataset.}
            \label{fig:data-over-written}
        \end{figure}

        A common adaptation the participants had to make to the generated code was removing lines that were already present in their notebook. This mostly included (1)~repeated library import statements---especially \pythoninline{import pandas as pd}---which were generated with every code response generated by ChatGPT, (2)~repeated data ingestion using the 
        \pythoninline{pandas.read_csv()} API, 
        or by hard-coding a sample of rows as a python dictionary while using the \pythoninline{pandas.DataFrame.from_dict()} API, and (3)~performing exploratory data analysis using \pythoninline{pandas.DataFrame.describe()} and other APIs. Accidentally executing this code sometimes led the participants' dataframes to be over-written by the sample values generated by ChatGPT (Figure \ref{fig:data-over-written}), causing them to re-run all notebook cells from the beginning (P1, P9).

    \subsubsection{Data and notebook management preferences} 
    \label{code-mgmt-preferences}
    We now discuss a few issues related to data and notebook management preferences:

    \smallskip
    
        \textbf{\textit{Managing generated code in computational notebooks.}} We observe that data scientists have implicit preferences for code management, aligned with common practices in the use of computational notebooks, such as having all import statements in a single cell at the beginning of the notebook, splitting functions and data transformations into cells to view intermediate results, interleaving of code with markdown cells to segment the notebook, and so on. 

        ChatGPT often generated big chunks of code involving multiple data transformations as a single code-block in its responses. Participants often broke-down such code chunks into meaningful cells (P2, P3, P5, P7, P8, P14). This helped them in isolating lines of code into steps, and execute them sequentially to observe any execution errors. This also enabled them to view the dataframe in between the data-transformation steps, while ensuring that the generated code matches their intent. Further, ChatGPT bundled library import statements with other data transformation/visualization code in the same block. This frequently required participants shifting the new import statements to the cells at the top of their notebook (P3, P5, P7, P12, P13).
        
        Data scientists have also been studied to commonly author exploratory code; or use the fluent programming pattern, composing multiple transformations into a chain \cite{Kery_VLHCC17, FluentInterface_Fowler_2005}. When participants mentioned the phrase ``write a function'' in their prompts, it led to generation of functions for data transformation tasks, with excessive number of parameters and intermediate steps, contrary to the typical exploratory and fluent patterns. P3 and P7 refactored such code generations to obtain chain of transformations for the target column.

         \smallskip
         
        \textbf{\textit{Data transformation and management.}} Participants often had preferences on how data transformation must be performed, such as---deciding to retain the parent column(s) after having derived newer column(s) (P3, P8, P9, P12), creating temporary dataframes for data transformation tasks (P12, P13), and deciding on appropriate techniques for data imputation (P6). Adaptations had to be made to the generated code to take these transformation preferences into account. For instance, when P6 prompted ChatGPT for T2 (split using delimiter), it generated an additional line of code to impute missing values in zipcode with \texttt{``NULL''}. P6 said, \emph{``No, no, I would not have preferred filling them with `NULL' because right now probably I still need to understand these columns more, so I may just remove null values afterwards,''} and subsequently removed the line of code.

    \subsubsection{Code validation}
    \label{code-validation}
        Participants emphasized on the importance of thoroughly validating every operation performed using LLM-generated code. P7 mentioned that validation is even more essential as \emph{``ChatGPT's confirmatory and affirmative language in responses---like `Definitely! Here's the code you need'---is extremely deceiving because it doesn't really know about my data.''}
        
        Though code validation took the least amount of time when compared to other activities (prompt construction and code adaptation) (Figure \ref{fig:time-spent-percentage}), each task required participants to verify the functioning of the generated code in different ways. In some cases, ChatGPT self-generated the code snippet(s) required to validate the data transformation. In other cases, the participants wrote additional code snippets, or manually inspected the data to validate changes. For instance, to ensure that \texttt{`INCIDENT\_DATETIME'} has been typecast to \texttt{pd.datetime} (T1), participants used \pythoninline{df.dtypes} and \pythoninline{df.columns}. To check for newly derived columns after splitting \texttt{`INCIDENT\_DESCRIPTION'} (T2), participants printed and manually inspected specific columns. Some participants also checked for missing values after using \pythoninline{df.value_counts()}. On the other hand, for T4 (heatmap plotting), visual verification of the generated plot would be sufficient to validate the code generated by ChatGPT.

\subsection{Strategies for Prompting and Alternate Resources}
\label{strategies}
    
    We now discuss strategies employed by data scientists to overcome communication and code-adaptation obstacles, along with scenarios where they prefer to leverage other resources (e.g., web-search, documentation, and previously written code).
    
    \subsubsection{Techniques for prompt construction}
    \label{placeholders}

        Alongside data context, participants also brainstormed prompting techniques that are relevant for ChatGPT to solve the problem. We observed frequent use of one-shot and few-shot prompting for T1 (datetime typecast) and T2 (split using delimiter) (P1--P4, P6, P8, P10, P12, P14); chain of thought prompting for T3 (feature selection), to elicit reasons behind \textit{why}, and \textit{on what basis} ChatGPT suggests features (P1, P3, P10); and making ChatGPT assume the role of a data scientist to receive expert suggestions, by prefixing each prompt with \emph{``You are a data scientist''} (P13).
        
        To avoid the frequent context-switching between prompt-writing (in the ChatGPT browser-window) and data context gathering (from the notebook/Excel), we observe that participants focused on creating a prompt skeleton while embedding ``placeholders'' for data context that must be fetched from the notebook (discussed in section \ref{context-sharing}), and pasted into the prompt (P1, P2, P5, P7, P8, P10, P14). For instance, P2 created placeholders (enclosed in square brackets), to replace them afterwards with the required column name and sample value:

    \begin{quote}
    \itshape
    I have a dataframe with a column called \textbf{[col name]}. This column has data separated by ;, here is an example: \textbf{[sample value]}. Can you split this column? (P2)
    \end{quote}
    
        This strategy was commonly used by participants to complete writing the natural-language query in one go, instead of their flow being disrupted by going back and forth between ChatGPT and their notebook to fetch relevant data context. Prompt drafts with placeholders also offered a low cost in switching out pieces of data context while assessing their importance for solving the problem at hand.

    \subsubsection{Scaffolding with domain expertise}
    \label{scaffolding-domain-expertise}
        Participants often expressed that ChatGPT lacks \emph{``any actual understanding of the data''} (discussed in sections \ref{assumptions} and \ref{misaligned-expectations}). To overcome these obstacles, we observe limited interactions where data scientists successfully guide the LLM with their domain expertise. P10 and P14 narrowed down ChatGPT's output space by iterative refinement of the data context included in their prompts. 
        
        To obtain meaningful visualizations (T4), P10 refined their prompt to only include names of \texttt{float} columns.

        \begin{quote}
        \itshape
        I would like to plot out response time vs:
        
        [List of `float' attribute names]
        
        Can you write python code that does this in seaborn? (P10)
        \end{quote}        
        
        Similarly, to prevent ChatGPT from suggesting timestamp-based features (T3), P14 provided only names of columns that do not contain the sub-strings `TIME' or `ID' in them.    

    \begin{quote}
    \itshape
    I want to build a GBM using sklearn. The X variables are [List of attribute names not containing sub-strings `TIME' or `ID'], target variable is ``RESPONSE\_TIME'' which is timedelta (P14)
    \end{quote}

    Surprisingly, we also observe an instance with P5, where ChatGPT asks a ``clarifying question'' in response to the user provided prompt:

        \begin{quote}
        \itshape
        P5: Consider this dataset to be the dataframe which we are going to use for the prediction of the response time: [Top 10 rows with data header]
        \end{quote}
        \begin{quote}
        \itshape
        ChatGPT: What is response time in this dataset and how is it calculated?
        \end{quote}
        \begin{quote}
        \itshape
        P5: Response time is calculated as below: FIRST\_ON\_SCENE\_DATETIME – INCIDENT\_DATETIME
        \end{quote}

        This interaction made P5 feel assured that ChatGPT \emph{``is able to understand the prompt''}, and \emph{``make sense of the data context''}. The clarifying question enabled P5 to provide context that they missed in their initial prompt, and allowed for complete intent expression instead of generating a hallucinated response. 

\subsubsection{Choosing alternate resources over ChatGPT}\label{alternatives}

There were several factors that lead to a preference to use alternative resources over ChatGPT:

\smallskip

\textbf{\textit{Data privacy concerns.}} Participants reflected on the importance of maintaining privacy for sensitive and personally identifiable information (PII). P2 was hesitant about using any confidential business data with ChatGPT and said, \emph{``Do you want me to use normal ChatGPT? And we are not using that for any sensitive data, right, because I don't want to share any private data with ChatGPT.''} Contemporary tools for data might inherently include data context in their prompt, to scaffold user queries. For this reason, participants expressed interest in controlling and reviewing any data context or information being passed on to the LLM. P3 expressed concerns over having their data become part of the model's training and being presented to a different ChatGPT user as a result to their queries. Participants were comfortable with using enterprise ChatGPT only in accordance with company policy (P2, P7).

\smallskip

        \textbf{\textit{Re-using previously written code.}} Data scientists mentioned that they periodically receive new batches of data, for which they must re-run analyses and visualization reports, and re-train models (P3, P5, P6, P7, P11). In such cases, data scientists are already familiar with the structure of the data, and they reuse previously written code for data pre-processing and cleaning. P11 mentioned, \emph{``We pretty much just copy and paste canned code for every project. A lot of variations can be created on the charts, but 80\% of it would be repeating.''} The need to author new code for pre-processing or analytics in such scenarios is rare.

\smallskip

        \textbf{\textit{Using web search and documentation.}} Participants described that using LLMs can normally help them save time over making a web search, such as when \emph{``asking for generic directions and steps, saving me the time to go through 4--5 blog posts on the web,''} and \emph{``asking for code in my specific context, which needn't be the exact issue discussed on Stack Overflow''} (P4, P11). In these cases, they find the LLM's responses to be highly contextualized to their data and previously applied steps, and \emph{``less noisy''} compared to web-pages. On the other hand, P7 found it easy to make a web search, requiring only \emph{``8--10 key words,''} whereas they must write \emph{``longer queries and be more specific with ChatGPT just because it allows me to do that''}. P11 also mentioned that the use of LLMs will \emph{``restrict me from learning new techniques or discovering a breadth of alternate approaches. Forums like Stack Overflow support these needs better.''}


\section{Survey Results}
\label{Section:CSurvey}
Our qualitative insights from the task-based study (Section~\ref{sec:observations}) identify numerous obstacles that data scientists face in conversing with ChatGPT, and subsequently, in acting on the generated responses. The survey results help us understand how well do these obstacles generalize to a diverse set of data science tasks, and to a broader community of data scientists. The survey respondents report experience with using LLMs for varied data science tasks spanning across project timelines, including synthetic data generation, data wrangling, pre-processing, data labelling, exploratory data analysis, insight finding and summarization, hypothesis generation, training various models, outlier detection, and generating plots. 

\begin{figure}[h]
    \centering
    \captionsetup{justification=centering}
    \includegraphics[width=\linewidth]{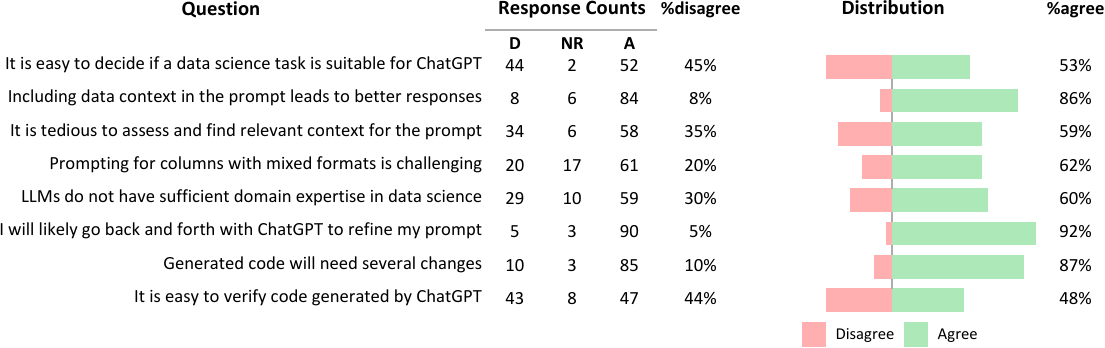}\
    \caption{Distribution of survey responses for `Agree'/`Disagree' questions. Response Counts: D (Disagree); NR (No Response); A (Agree). }
    \label{fig:survey-results}
\end{figure}

Figure~\ref{fig:survey-results} shows the distribution of responses for each `Agree'/`Disagree' question. Results indicate that data scientists find sharing of relevant data context to be important (86\%), as well as tedious (59\%), echoing findings from Section \ref{context-sharing}. 62\% data scientists agreed that prompting ChatGPT for columns with mixed formats is challenging, reflecting the issues discussed in Section \ref{assumptions}. These challenges lead to communication breakdowns between data scientists and ChatGPT---causing them to make repeated prompt refinements (92\%) (Section~\ref{sec:prompt_behavior}). Respondents also indicate that LLMs do not have sufficient domain expertise in data science (60\%) (Section \ref{misaligned-expectations}), and that generated code requires several modifications before it can be used in their notebooks (87\%) (Section \ref{obstacles-adapt-code}). Lastly, code validation respondents have a split opinion on whether it is easy to verify code generated by ChatGPT (48\%) (Section \ref{code-validation}).

We manually inspected responses to look for any other challenges encountered by data scientists in conversing with LLMs for data science tasks. Respondents re-emphasized difficulties in \emph{``forming well scoped prompts''} and deciding \textit{what} context to include. They reflected on the need for repeated prompt refinements due to ChatGPT's lack of domain expertise. A few responses brought up the \emph{``instability''} in ChatGPT's responses, wherein it generates drastically different answers despite sending the same prompt. We did not encounter any new obstacles in response to this question.


\section{Design Recommendations}\label{sec:design}

In this section, we discuss the implications of our findings and identify the ways in which they could be adapted to various data-science environments and to broader contexts. 

\subsection{Recommendation I---Provide preemptive and fluid context when interacting with AI assistants}

From our studies, we found that participants spent a significant amount of time constructing prompts (Section ~\ref{sec:prompt_behavior}) and gathering and expressing their context (Section~\ref{context-sharing}) to ChatGPT. Nearly 45\% of the prompts included portions of data that they either manually entered, or wrote additional code snippets to obtain; however, participants did not enjoy being \emph{``slowed-down''} in fetching this data context for every interaction. Thus, data scientists may need help in providing context about their environments, and we should support mechanisms to either preemptively provide implicit context, or provide fluid interactions that allow data scientists to easily refer to fragments of their work in conversations with AI assistants.

\begin{figure}[h]
        \centering
        \includegraphics[width=\linewidth*7/10]{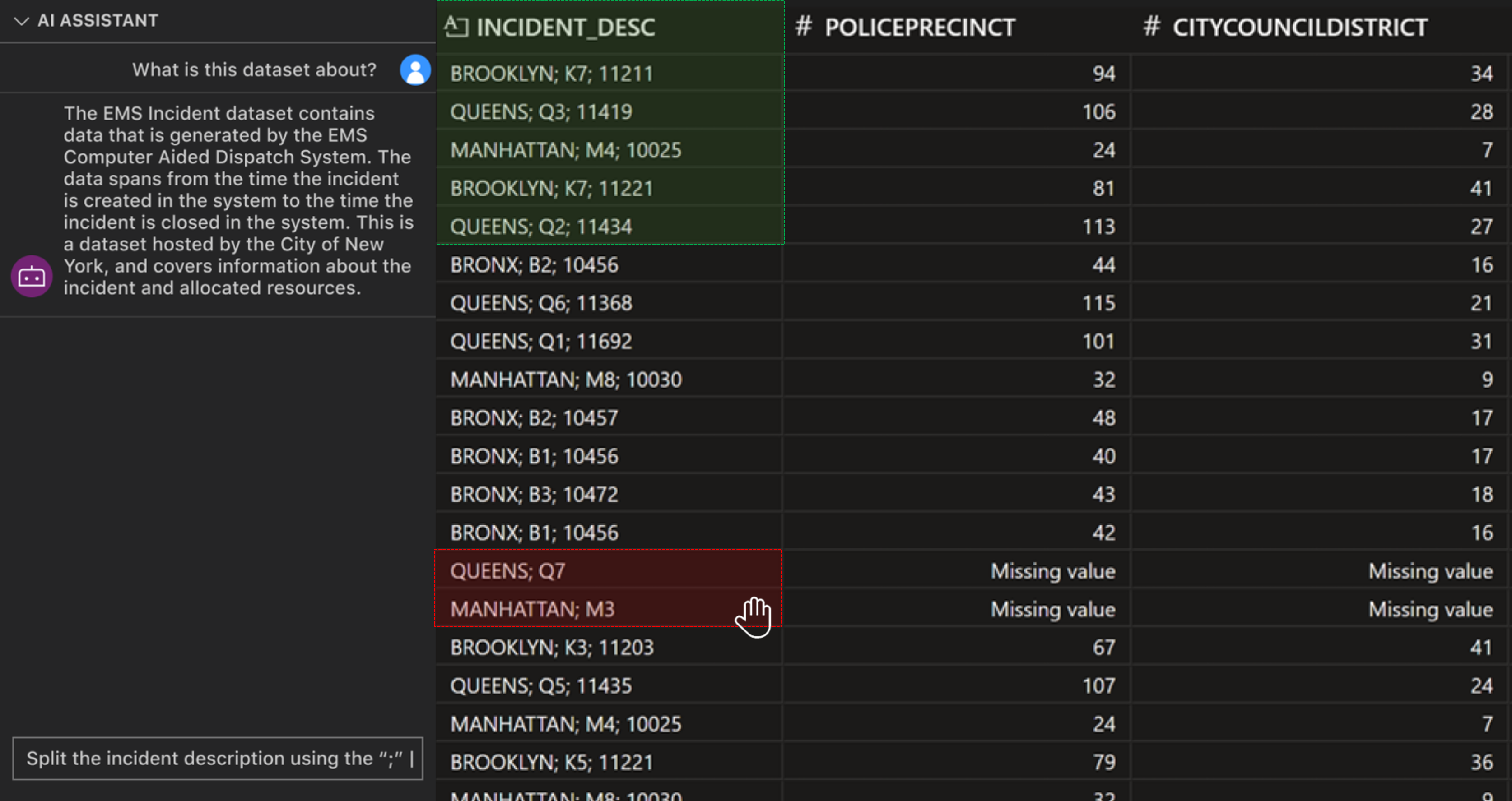}\
        \caption{Design concept---Using data brushing to mark examples (selected in green), and inconsistent entries (selected in red).}
        \label{fig:data-brushes}
\end{figure}

In data visualization, classic techniques such as \emph{brushing}~\cite{Becker:1987}, enable a user to select and highlight specific data points, allowing the user to use those data points as a filter, or as a point of interest for further analysis. Similarly, providing mechanisms for supporting \emph{data brushing} in data-science environments would allow users share their context with agents more directly. For example, in T2, participants had to split text into three columns. However, some rows had inconsistent formatting making the operation suggested by ChatGPT error-prone. With data brushing (Figure~\ref{fig:data-brushes}), participants could mark an example cell with a problematic entry and ask an AI assistant to help split the rows, while dealing with outliers such as the marked cell. Similarly, allowing inline code snippets that are expanded, or referring to entities within the data (\texttt{@INCIDENT\_RESPONSE)}, allow less tedious and more fluid expressions of context when prompt writing.

Despite these aids, some context, especially providing a summary and schema of the data would be tiresome. Furthermore, users had no clear understanding of what information should be provided, and even for researchers, this question is still being explored~\cite{chen2023large}. However, recall Grice's maxim of relevance, which states that a successful conversation requires not too little or too much information be provided~\cite{Grice1991Book}. A simple intermediate measure would be to provide a preemptive summaries of data characteristics. This would help reduce the chances that ChatGPT makes assumptions about data format, which may lead to incorrect code or longer conversations (Section \ref{assumptions}).

\subsection{Recommendation II---Provide inquisitive feedback loops and validation-aware operations}
For transactional tasks, such as T1 (datetime typecast), ChatGPT often provided satisfactory responses. Unfortunately, open-ended tasks, such as T3 (feature selection), required a more involved conversation with ChatGPT.
Initial responses from ChatGPT were often ``\emph{excessively long}'' and were not particularly helpful for the task, likely because it violates the fundamental HCI principle of progressive disclosure~\cite{Nielsen2006} and overwhelms the user with details.
As a result, participants were uncertain how much value would be gained by continuing the conversation or what further information would help (Section~\ref{misaligned-expectations}). In other cases, participants had to make multiple clarifications and carry out  editing operations across several iterations in their data-science environments. This task fragmentation often lead to errors, for example, because ChatGPT always included code to ingest the dataframe---regardless of the stage of conversation---some participants accidentally executed it, leading to their data being over-written (Section~\ref{repeated-steps}).

One mechanism to help participants have more productive conversations with AI assistants is through \emph{inquisitive feedback loops}, where the system guides the user in expressing what they need by proactively asking the user a series of questions, rather than waiting for the user to know when they need help and how to get it~\cite{Henley:2021}.
Moreover, participants in our study could have been aided just by ChatGPT asking clarifying questions to resolve ambiguities instead of making assumptions.
For example, one participant (P9) was able to get more useful advice from ChatGPT, after suggesting that they could reframe the problem of feature selection: ``\emph{I think I am more interested in finding the feature importance so I'll prompt it for something like a random forest}''. In this case, a user could be prompted with questions that helped the user think through which high-level strategy they wanted to use before giving all the details.

Tools for supporting \emph{refactoring}, behavioral-preserving transformations, can be one approach for assisting data scientists who must make repeated edits of similar nature, or adapt generated code snippets to their local notebook contexts (Section~\ref{obstacles-adapt-code}). For instance, \textsc{GhostFactor}~\cite{Ge:2014} is a light-weight program analysis tool with refactoring detection algorithms that can identify incomplete changes to code. Simple approaches can be used for enabling automated post-processing and human-in-the-loop refactoring of generated code snippets based on observed nature of edits---deletion of repeated code, 
notebook and data management preferences, and validation. 

\subsection{Recommendation III---Provide transparency about shared context and domain expertise slots}
\citet{Barke_OOPSLA23} posit that lack of transparency about the shared context, and lack of control in refining the context selections can leave users in a confused state. Our participants echo a similar consideration, wherein they expressed the urge to \emph{``get to the first response as fast as possible, but have the ability to look at what context was provided and edit it if I need to refine my prompt''} (P3, P4, P7, P12). Furthermore, participants also wanted to be able to provide more context that reflected their personal expertise and experience (Section~\ref{scaffolding-domain-expertise}).

Even with automated context management or prompt augmentation, it is important to consider how that context is ultimately shared. For example, 
by having sufficient transparency into the shared context can also enable data scientists to ensure that sensitive data does not leak into the prompt.
Data-science environments interacting with AI should consider mechanisms for 
ensuring a two-way knowledge transfer between data scientists and AI. For example, an AI assistant chat interface could integrate a \emph{context panel}, a dedicated, editable grid that mirrors the AI assistant's and user's assumptions about data and tasks.

Finally, users should be able to inspect and modify the context to provide their domain expertise or reflect their analysis style (e.g., always perform data scaling and impute missing values prior to training a model). One step in this direction is the recent addition to ChatGPT allowing for custom instructions\footnote{\url{https://openai.com/blog/custom-instructions-for-chatgpt}}.


\section{Threats to validity}

All study participants had prior experience with using ChatGPT. However, most participants (9 \textit{of} \StudyParticipants{}) had not used LLMs for solving data-science tasks. If they had been exposed to tasks of similar nature prior to the study session, their interactions with ChatGPT could have been different. We took this into consideration to ensure the quality of responses in the confirmatory survey, and asked all respondents to share at least one scenario where they used an LLM-powered tool for a data-science task. This enabled us to filter out respondents who did not have any prior experiences to provide informed responses. 

We selected a diverse set of data-science tasks representative of data pre-processing, cleaning, feature selection, and plotting. However, the study duration limited the breadth of tasks that participants could have performed, such as data discovery, capture, and machine learning~\cite{Piorkowski_CHI19}. Nevertheless, our survey responses (Section~\ref{Section:CSurvey}), from a broader community of data scientists with diverse experiences, confirm the generalizability of our findings. Future work can involve empirical studies that measure the performance of LLMs on assorted data-science tasks with datasets that have not been ``seen'' by the LLMs during pre-training. We presented the data-science tasks to all participants through verbal descriptions of the objectives. This may have inevitably led to priming, wherein the participants' formulation of prompts could have been influenced by our description of the task.

Browser-based version of ChatGPT version was updated as we conducted our task-based studies. P1--P9 used the March 14, 2023 version of ChatGPT, and P10--P14 used the March 23, 2023 version. This might have had a slight impact on study findings, though we did not observe any noticeable shifts in the nature of responses generated by ChatGPT for the study tasks. Lastly, we encouraged participants to think aloud as they solved study tasks, which could have added to the time taken in performing different activities. Our video annotations take this into consideration to only include timestamps where participants are actively working with the data, or with ChatGPT.


\section{Related Work}

AI-assisted tools for data cleaning suggestions---such as Trifacta\footnote{\url{https://docs.trifacta.com/display/SS/Overview+of+Predictive+Transformation}} \cite{Guo_UIST11, Wrangler_CHI11}, Data Wrangler\footnote{\url{https://devblogs.microsoft.com/python/data-wrangler-release/}}, CoWrangler~\cite{CoWrangler_SIGMOD23}, AWS Glue DataBrew\footnote{\url{https://docs.trifacta.com/display/SS/Overview+of+Predictive+Transformation}}, Salesforce Einstein Discovery\footnote{\url{https://help.salesforce.com/s/articleView?id=sf.bi_edd_prep_terminology.htm}}, AutoPandas~\cite{Rohan_OOPSLA19}, Auto-Suggest~\cite{Yeye_SIGMOD20}---have existed, but LLMs bring forth new opportunities for enhancing these tools further.
More recently, several commercial and open-source tools have emerged---DataChat AI\footnote{\url{https://datachat.ai/}}, 
Anaconda Assistant\footnote{\url{https://www.anaconda.com/blog/anaconda-assistant-launches-to-bring-instant-data-analysis-code-generation-and-insights-to-users}}, 
Databricks Assistant\footnote{\url{https://www.databricks.com/blog/introducing-databricks-assistant}}, 
and Jupyter AI\footnote{\url{https://jupyter-ai.readthedocs.io/}}---which enable data scientists to access AI-powered chat assistants within their notebook (such as Anaconda, Databricks, and Jupyter notebooks), and other alternate data-science environments (e.g., DataChat AI, SQL and file editors for Databricks, etc.). The semantic abilities of LLMs, coupled with a chat interface, provides users with the agency to converse about their data, ask follow-up questions, and receive highly contextualized responses.

McNutt et al.~\cite{McNutt_CHI23} studied, through an interview study, the design space of code assistants in computational notebooks, but their focus is on the design space as they do not focus on data-specific issues that data scientists face while communicating their needs to the AI assistants (e.g., communicating the data context or domain knowledge). A number of recent work conducted studies to understand the use of LLMs by software engineers and programmers~\cite{BAMyers_2023, Barke_OOPSLA23, Sarkar_PPIG22, Jayagopal_UIST22, Priyan_ICSE2022, Priyan_CHILBW2022, kaddour2023challenges}. However, these mostly focus on the general understanding of AI-powered code assistants and do not particularly focus on data-science tasks, where communicating the data at hand to the assistant is a big challenge.

There has been a recent line of work that try to make use of LLMs to automate data-science tasks and evaluate performance of LLMs on data-science tasks~\cite{Numeracy_2023, FeatureEngg_2023, Polozov_2022_NL2Code_Notebooks}. This is a promising, but orthogonal, area of research to ours as we focus on human-in-the-loop data science where AI-assistants and human data scientists work as a team.
Similarly, there are work on designing and evaluating tools for end users~\cite{Gordon_CHI23, Sruti_IUI22, ColDeco_VLHCC23}, which is complementary to our work.

Most relevant to our work is the study by Liu et al.~\cite{Gordon_CHI23} on how users of a spreadsheet software articulate their question to an AI-powered tool and how a system they designed can better support this process.
Our work is differentiated by (1)~examining the fundamental interactions with a general AI tool like ChatGPT rather than a domain-specific AI tool embedded in a system, (2)~we studied professional data scientists rather than non-programmers, and (3)~they were specifically investigating the effect of co-editing a prompt by splitting it into natural language steps.


\section{Conclusion and Future Work}

The recent rise of AI-powered chat assistants gives hope of greatly increasing the productivity of data scientists as they are engaged in tedious data-wrangling tasks.
To understand the fundamental obstacles, needs, and design opportunities of data scientists using AI assistants, we conducted mixed-method studies involving a task-based study, interviews, and a survey with professional data scientists.
We found that participants faced two key sets of barriers: when communicating with ChatGPT and then when adapting ChatGPT's response to aid their specific situation.
For example, participants spent considerable time collecting and formatting information to provide to ChatGPT, such as relevant regions of their dataset or descriptive statistics.
Furthermore, ChatGPT often made assumptions unbeknownst to the participant, thus leading to bugs in the code and additional steps to remedy.
When participants did get a useful response from ChatGPT, it required considerable effort to adapt the code to work with the rest of their code and to follow their style.
We provide key design recommendations for tools based on our study, such as the need to provide inquisitive feedback loops and validate-aware operations.
As a next step, we intend to operationalize these design guidelines into a working tool and assess its efficacy in alleviating the workflow challenges that data scientists encountered in our studies.
In conclusion, our work aims to function as a significant catalyst in democratizing data science via AI-powered chat assistants.


\bibliographystyle{ACM-Reference-Format}
\bibliography{bibliography}

\end{document}